\documentclass[pre,twocolumn,english,showpacs,aps,prb,a4paper,floatfix]{revtex4}
\usepackage[T1]{fontenc}
\usepackage{color}
\usepackage[latin1]{inputenc}
\usepackage{graphicx}
\usepackage{epsfig}
\begin{document}
\title{Effect of polydispersity, bimodality and aspect ratio 
on the phase behavior of colloidal platelet suspensions}

\author{Yuri Mart\'{\i}nez-Rat\'on}
\affiliation{Grupo Interdisciplinar de Sistemas Complejos (GISC),
Departamento de Matem\'{a}ticas,Escuela Polit\'{e}cnica Superior,
Universidad Carlos III de Madrid, Avenida de la Universidad 30, E--28911, Legan\'{e}s, Madrid, Spain}

\author{Enrique Velasco}
\affiliation{Departamento de F\'{\i}sica Te\'orica de la Materia Condensada
and Instituto de Ciencia de Materiales Nicol\'as Cabrera,
Universidad Aut\'onoma de Madrid, E-28049 Madrid, Spain}

\date{\today}

\begin{abstract}
We use a Fundamental-Measure density functional for hard board-like polydisperse particles, in the restricted-orientation
approximation, to explain the phase behaviour  
of platelet colloidal suspensions studied in recent experiments. In particular, we focus our attention on the behavior
of the total packing fraction of the mixture, $\eta$, in the region of two-phase  
isotropic-nematic coexistence as a function of mean aspect ratio, polydispersity and fraction of total volume 
$\gamma$ occupied by the nematic phase. 
In our model, platelets are polydisperse in the square section, of 
side length $\sigma$, but have constant thickness $L$ (and aspect ratio $\kappa\equiv L/\langle\sigma\rangle<1$, 
with $\langle\sigma\rangle$ the mean side length). 
Good agreement between our theory and recent experiments is obtained by mapping the real system onto an effective one, with excluded volume 
interactions but with thicker particles (due to the presence of long-ranged repulsive interactions 
between platelets). The effect of polydispersity in both shape and particle size has been taken into account 
by using a size distribution function with an effective mean-square deviation that depends on both polydispersities. 
We also show that the bimodality of the size distribution function is required to correctly describe the huge 
two-phase coexistence gap and the nonlinearity of the function $\gamma(\eta)$, two important features that these colloidal 
suspensions exhibit. 
\end{abstract}

\pacs{61.30.Cz, 61.30.Hn, 61.20.Gy}

\maketitle

\section{Introduction}
Colloidal suspensions of mineral or viral anisotropic particles interacting via short-ranged repulsive forces 
exhibit a phase behavior with entropy-driven phase transitions between their liquid-crystalline phases. The nature 
of these phases strongly depends on particle geometry. The rod geometry in mineral or viral 
particles favours the formation of uniform phases, i.e. isotropic (I) and nematic (N) phases \cite{Varga}, and also
of the layered smectic (S) phase \cite{Maeda,Dogic,Purdy}. In the case of the plate geometry, the I and N phases are usually 
stabilized at low particle concentration \cite{Kooij2}, and the N phase requires a relatively high
aspect ratio (and thus particle anisotropy) \cite{Kooij2}. In addition, as the volume fraction increases, there may 
appear a transition to the columnar (C) phase \cite{Petukhov,Byelov}. 

The plate geometry is more versatile as regards the type of liquid crystalline phases it may induce \cite{Kooij1}.
Recently it {has} been shown that colloidal suspensions of some plate-like, mineral 
charged particles, can also stabilize the smectic phase \cite{Sue,Kleshchanok}. The colloidal particles are usually polydisperse in their sizes (diameter and 
thickness) and shapes (rod or plate geometry or different particle cross-sections), and it was found
that polydispersity causes phase behavior in these systems to be much more complex due to phenomena such as 
size segregation, fractionation and multiple phase coexistence \cite{Kooij3}. For example,
polydisperse rod-plate mixtures exhibit demixing between I and different N phases (with the former populated by particles 
with less anisotropy), and also up to four coexisting phases  
(some of them nonuniform) may exist at high densities \cite{Kooij3}. No trace of biaxial N phases was found in these mixtures. 
However, recent experiments on board-like colloidal particles could find this elusive phase \cite{vandenPol}. 
Also I-N coexistence with density inversion has been observed \cite{Wensink,Verhoeff} (with the I phase being the densest phase).  

The theoretical modeling of these kind of mixtures turns out to be a difficult task. Density functional theory (DFT), which
is based on the local density distribution,  
has been successful in the description of bulk and interfacial phase behavior of hard spheres and other
fluids of anisotropic particles \cite{Tarazona}, but it becomes hard to implement for the case of polydisperse mixtures. 
This is due to the huge number of degrees of freedom on which the local density now depends:
not only on spatial and (for anisotropic particles) orientational coordinates, but also on a number of polydispersity   
variables. Despite the increased difficulty, some theoretical calculations on polydisperse mixtures 
of freely rotating hard rods in the Onsager approximation 
have been performed \cite{Speranza1,Speranza2}. These calculations confirm the phase behavior found in experiments as regards 
the broadening of the I-N coexisting gap and the size-fractionation phenomenon. In this respect, it would be interesting to 
extend the recently proposed Fundamental-Measure DFT for freely rotating hard disks \cite{Schmidt} to the calculation of 
phase behavior in polydisperse platelets. 

Monte Carlo simulations on polydisperse infinitely thin hard-platelet 
fluids have also been carried out \cite{Bates}. These results again show the dramatic effect of polydispersity
on phase behavior in hard-platelet fluids (mainly segregation driven by particle size), as compared to that in one-component 
fluids, already simulated in the 90s for the cut-sphere geometry \cite{Veerman}. In the latter work
the bulk phase diagram was traced. For aspect ratios $\kappa\equiv L/\sigma<1$ the N and, at higher densities the C, phases
were found to be stable (here $L$ is the thickness and $\sigma$ the particle diameter). In the range 
$0.1\alt \kappa\alt 0.2$ the C phase gives rise {to the I and eventually for 
$\kappa \sim 0.2$ to the cubatic phase}, and finally 
for $\kappa\agt 0.3$ only the I and the solid phase are stable. Similar phase behavior was found in Monte Carlo simulations of 
hard oblate spherocylinders \cite{Marechal}, where two different crystals are stable (tilted for $\kappa\alt 0.45$ 
and aligned for $\kappa\agt 0.45$); the cubatic phase is always unstable. The practical difficulties in implementing
DFT calculations associated with polydispersity can be circumvented by considering discrete particle orientations, 
as in the Zwanzig model where the main axes of the particles, one for uniaxial and two for biaxial geometries, 
point along one of the three Cartesian axes \cite{Zwanzig}. In the framework of this model the phase diagrams 
of polydisperse hard rods \cite{Clarke} and rod-plate mixtures \cite{Yuri1} have been calculated. 

Within the same approach, the effect of polydispersity on the stability of 
the biaxial N phase of hard board-like biaxial particles \cite{Roij} has recently been studied.
The phase diagram of the one-component limit of this fluid 
was recently obtained \cite{Yuri2} using a DFT based on Fundamental-Measure theory for hard biaxial 
parallelepipeds \cite{Cuesta}. Finally, the same model has been applied to the study of interfacial properties of 
binary mixtures confined by external potentials \cite{Harnau}.

Recently a systematic experimental study of the phase behaviour of polydisperse platelets in suspension
has been presented \cite{chinorri}. {Particles were synthesized by {
hydrothermal methods}, 
and further exfoliation (through chemical treatment with TBA molecules
the pristine Zirconium-Phosphate (ZrP) crystals are delaminated into single layers) 
\cite{chinorri}}. The novel feature of this fluid is that polydispersity is in 
the platelet diameter, with a strictly constant thickness. Also the shape of the particle cross sections is polydisperse,
with most particles having hexagonal geometry. The work was focused on the study of the I-N 
transition for different polydispersities and mean aspect ratios. The aim of the present article is to theoretically 
understand the results presented in Ref. \cite{chinorri}. The experimental results and the conclusions we obtain from our
theoretical model can be summarized as follows: 

(i) Highly polydisperse platelet mixtures exhibit a huge I-N coexistence gap which cannot 
be explained by simply assuming a unimodal distribution function for particle diameters. However, if one considers a {\it bimodal}    
distribution function, the coexistence gap can be explained via a demixing mechanism. We remark that bimodality may {\it not} be 
apparent through a direct visual inspection of the bimodal distribution function. Note that the effect of bimodality in 
platelet thickness (not diameter) on phase behavior has already been studied in Ref. \cite{Verhoeff}, but in that case
the bimodality in the size distribution function is clearly seen from the size histogram, which is not the case in Ref. 
\cite{chinorri}.  

(ii) The repulsive character of colloidal platelet interactions in the experimental system is incorporated through  
an effective platelet thickness $L_{\rm eff}$ which is much higher than the thickness of the real platelets. The effective
thickness is chosen to guarantee a reasonable description of experimental data by the theoretical model. 

(iii) In order to account for the shape polydispersity and to adequately describe the experimental findings, the polydispersity 
coefficient (mean square deviation of the size distribution function) should be taken much higher 
than that given in Ref. \cite{chinorri}. In the present study we have used a DFT for hard board-like polydisperse particles 
with square cross sections of width $\sigma$ and constant thickness $L$, in the oblate particle regime $\kappa=L/\sigma<1$. 
The DFT is based on the FMT for the hard-parallelepipeds in the restricted-orientation approximation \cite{Cuesta}. 

The article is organized as follows. In Sec. \ref{theory} we present the theoretical model, making special emphasis on the 
implementation of the I-N coexistence calculations (Sec. \ref{coexistence}), the size distribution function used 
to model polydispersity (Sec. \ref{bimodal}) and the effect of particle shape polydispersity on the 
effective size polydispersity of the mixture (Sec. \ref{shape}). Sec. \ref{results} presents the results, and is divided 
in two sections: Sec. \ref{uni}, which is devoted to the phase behavior of the mixture assuming a unimodal size distribution 
for diameters, and Sec. \ref{bi}, which presents results obtained with bimodal distributions.  
Finally we draw some conclusions in Sec. \ref{conclusions}.

\section{Theory }
\label{theory}
The theory we use is based on a density functional for hard board-like particles, formulated in the restricted orientation approximation
(the so-called Zwanzig model). Particles are polydisperse in the side length $\sigma$ of the square section, but their thickness $L$ is 
fixed. The mean aspect ratio $\kappa\equiv L/\langle\sigma\rangle$ (with $\langle\sigma\rangle$ the mean side length) is less than
unity, $\kappa<1$, so that we are in the oblate-particle regime. The main quantities that describe our model 
are the set of density distribution functions $\rho_{\nu}(\sigma)$, where $\nu=\{x,y,z\}$ refers to particles with their main 
axis parallel to the $\nu$ Cartesian axis; each of the three orientations can be considered to correspond to a different species, and the
fluid can therefore be treated as a three-species mixture. In the following sections we describe the theoretical formalism we have used 
to calculate the isotropic (I)-nematic (N) coexistence in polydisperse mixtures that possess unimodal/bimodal size distributions.    

\subsection{Coexistence calculations}
\label{coexistence}

If a fraction $\gamma$ of the total volume of the system $V$ is occupied by a nematic (N) phase 
in coexistence with the isotropic (I) phase, then the density distribution functions $\rho_{\nu,s}(\sigma)$ of 
the two phases, $s=\{\rm{I,N}\}$ should fulfill the lever rule (conservation of the total number of particles):
\begin{eqnarray}
\rho_0(\sigma)=\gamma \sum_{\nu} \rho_{\nu,{\rm N}}(\sigma)+
(1-\gamma)\sum_{\nu}\rho_{\nu,\rm I}(\sigma).
\label{const}
\end{eqnarray}
The density distributions of all species are the same for the I phase, $\rho_{x,{\rm I}}(\sigma)=\rho_{y,{\rm I}}(\sigma)=
\rho_{z,{\rm I}}(\sigma)\equiv\rho_{\rm I}(\sigma)$,
while for the uniaxial N phase we have $\rho_{x,{\rm N}}(\sigma)=\rho_{y,{\rm N}}(\sigma)\equiv\rho_{\perp,{\rm N}}(\sigma)$ and 
$\rho_{z,{\rm N}}(\sigma)\equiv \rho_{\parallel,{\rm N}}(\sigma)$; here we take the nematic director to be parallel to the $z$-axis. 
The density distribution function 
$\rho_0(\sigma)=\rho_0h(\sigma)$ ({\it parent} distribution function) 
is a product of the mean parent number density $\rho_0$ and the size distribution function $h(\sigma)$ which fulfills the normalization 
condition $\int d\sigma h(\sigma)=1$. {Note that $h(\sigma)$ has units of $[\rm Length]^{-1}$. As $\rho_0$ is the number density 
having units of $[\rm Length]^{-3}$ the distribution functions $\rho_0(\sigma)$ and $\rho_{\nu,s}(\sigma)$ have units of 
$[\rm Lenght]^{-4}$}. 

For the whole system I$+$N (with $0\le\gamma\le 1$), we define a Lagrange functional from the free-energy density in reduced thermal units, 
$\tilde{\Phi}=\beta F/V$ (with $F$ the free energy, $\beta=1/kT$, $k$ Boltzmann constant and $T$ the temperature), as  
\begin{eqnarray}
&&\tilde{\Phi}\left[\{\rho_{\nu,s}\}\right]=\gamma \Phi\left[\{\rho_{\nu,{\rm N}}\}\right]+(1-\gamma) 
\Phi\left[\{\rho_{\nu,\rm I}\}\right]\nonumber\\
&&+\int d\sigma {\mu_0(\sigma)} 
\left[\rho_0(\sigma)-\gamma \sum_{\nu} \rho_{\nu,{\rm N}}(\sigma)\right.\nonumber\\
&&\left.-(1-\gamma)\sum_{\nu}\rho_{\nu,\rm I}(\sigma)\right],
\label{total}
\end{eqnarray}
{where $\mu_0(\sigma)$ are the Lagrange multipliers 
that guarantee the constraints (\ref{const}). Note that $\mu_0(\sigma)$ is just the scaled with $\beta$ chemical 
potential of species $\nu$ of width $\sigma$ 
in each one of the coexisting phases, i.e. 
$\beta \mu_{\nu,s}(\sigma)=\mu_0(\sigma)$, $\forall$ $\nu=x,y,z$ and $s=\rm{I,N}$}.  
$\Phi$ is split in ideal
\begin{eqnarray}
\Phi_{\rm id}\left[\{\rho_{\nu,s}\}\right]=\int d\sigma \sum_{\nu} \rho_{\nu,s}(\sigma) 
\left[\ln \rho_{\nu,s}(\sigma)-1\right],
\end{eqnarray}
and excess $\Phi_{\rm exc}$ parts. In our treatment, we obtain the latter from Fundamental-Measure Theory (FMT). 
In the FMT formalism $\Phi_{\rm exc}(\{\rho_{\nu,s}^{\alpha}\})$ 
is a function of a finite number of moments of the distribution function. 
{The latter are defined as 
\begin{eqnarray}
\rho_{\nu,s}^{(\alpha)}=\int d\sigma \rho_{\nu,s}(\sigma)\sigma^{\alpha}, \quad, \alpha=0,1,2.
\label{mm}
\end{eqnarray}
The expression for the 
function $\Phi_{\rm exc}(\{\rho_{\nu,s}^{\alpha}\})$ [see Eq. (\ref{laenergia})] 
is obtained in the Appendix 
from the scaled particle theory (SPT), the uniform limit of the fundamental measure free-energy density functional 
\cite{Cuesta}}.  
The constrained functional minimization of  
(\ref{total}) with respect to $\rho_{\nu,s}(\sigma)$, together with the definition (\ref{mm}), 
provide   
the moments at equilibrium {\cite{Clarke}}:
\begin{eqnarray}
\rho_{\nu,s}^{(\alpha)}=\rho_0\int d\sigma
\frac{h(\sigma) \sigma^{\alpha}e^{-\beta\mu_{\nu,s}^{(\rm exc)}(\sigma)}}
{\displaystyle\gamma \sum_{\nu}e^{-\beta \mu_{\nu,\rm N}^{({\rm exc})}(\sigma)}+
3(1-\gamma)e^{-\beta \mu^{({\rm exc})}_{\rm I}(\sigma)}}, \nonumber\\
\label{auto}
\end{eqnarray}
where 
\begin{eqnarray}
\beta \mu_{\nu,s}^{(\rm exc)}(\sigma)=\sum_{\alpha=0}^2
\frac{\partial \Phi_{\rm exc}}{\partial \rho_{\nu,s}^{(\alpha)}}\sigma^{\alpha},
\end{eqnarray}
are the excess part of the chemical potential of species $\nu$ of width $\sigma$ in the phase 
$s$.
Note that this is  
a quadratic polynomial in $\sigma$ whose coefficients are in turn functions of 
the moments {$\rho_{\nu,s}^{(\alpha)}$} {\cite{Sollich3}}. We use the notation 
{$\rho^{(\alpha)}_{\rm{I}}\equiv \rho^{(\alpha)}_{\nu,\rm{I}}$ 
and $\mu^{(\rm{exc})}_{\rm I}(\sigma)\equiv \mu_{\nu,\rm I}^{(\rm{exc})}(\sigma)$ $\forall \nu$}. 
Thus the set of nine equations (\ref{auto}) (which guarantee the 
equality of chemical potentials of all species in both phases) are solved self-consistently 
for the moments {$\rho_{\nu,s}^{(\alpha)}$}, while the other 
quantity to be determined, $\rho_0$, is found by imposing
the condition of mechanical equilibrium, i.e. the equality of pressures [see Eq. (\ref{lapresion})] 
in both phases:
\begin{eqnarray}
P_{\rm I}\left(\rho_0,\{{ \rho_{\rm I}^{(\alpha)}}\}\right)=
P_{\rm N}\left(\rho_0,\{{ \rho_{\nu,{\rm N}}^{(\alpha)}}\}\right).
\end{eqnarray}  
The fluid pressure can be found from (\ref{lapresion}). 
The cloud-I--shadow-N coexistence, corresponding to a situation where the system volume is occupied by the I phase except for
a coexisting, infinitesimal amount of the N phase, is obtained by taking $\gamma=0$ in the expressions above. The case of the
shadow-I--cloud-N coexistence, which corresponds to the opposite case (i.e. the N phase occupying the whole system volume 
but in coexistence with an infinitesimal amount of the I phase), is obtained by taking $\gamma=1$. 

\subsection{Length distribution function}
\label{bimodal}

In the present study we choose a probability distribution function $h(\sigma)$ which is, in the 
general case, bimodal:
\begin{eqnarray}
\displaystyle{h(\sigma)=\frac{x}{\sigma_1}h_0\left(\frac{\sigma}{\sigma_1}\right)+
\frac{(1-x)}{\sigma_2}h_0\left(\frac{\sigma}{\sigma_2}\right)}.
\label{h_sigma}
\end{eqnarray}
Here $x$ is the molar fraction when the fluid is strictly a binary mixture; otherwise $x$ can be regarded as a 
parameter which controls the relative heights of the two maxima, located at $\sigma_1$ and $\sigma_2$. 
The function ${ h_0(u)}$ 
is selected to be

\begin{eqnarray}
h_0(u)=C u^{\nu}e^{-\Lambda u^p}, 
\end{eqnarray}
where the constants $C$ and $\Lambda$ are calculated from the normalization conditions
{$\int_0^{\infty} d u h_0(u)=
\int_0^{\infty} d u u h_0(u)=1$}. Thus we find 
\begin{eqnarray}
\Lambda^{1/p}=\frac{\Gamma\left[(\nu+2)/p\right]}{\Gamma\left[(\nu+1)/p\right]},\quad 
C=\frac{p\Lambda^{(\nu+1)/p}}{\Gamma\left[(\nu+1)/p\right]},
\end{eqnarray} 
with $\Gamma(x)$ the Gamma function. 
For fixed $p$ the parameter $\nu$ controls the polydispersity, while $p$ controls the decay of 
the distribution at large $\sigma$ (note that for $p=1$ a Schultz distribution is obtained). 
All these definitions guarantee the normalization  
$\int_0^{\infty} d\sigma h(\sigma)=1$. For the first moment we find $\int_0^{\infty} d\sigma\sigma 
h(\sigma)=x\sigma_1+(1-x)\sigma_2$. 
The one-component limit is recovered by setting $x=1$ and consequently $\langle \sigma\rangle=\sigma_1$.
Defining the polydisperse coefficient for {$h_0(u)$}  
as the mean-square deviation,  
{$\Delta_0=\sqrt{\langle u^2\rangle_0/\langle u\rangle_0^2-1}$}, where 
{$\langle u^{\alpha}\rangle_0\equiv\int_0^{\infty} d u  u^{\alpha} h_0(u)$}, 
we find that the 
polydispersity coefficient for the bimodal distribution function $h(\sigma)$, i.e.
$\Delta=\sqrt{\langle\sigma^2\rangle/\langle\sigma\rangle^2-1}$ [with 
$\langle\sigma^{\alpha}\rangle\equiv \int_0^{\infty} d\sigma \sigma^{\alpha} h(\sigma)$] results in
\begin{eqnarray}
\Delta=\frac{1}{\overline{\sigma}}\sqrt{\overline{\sigma^2}\Delta_0^2+x(1-x)(\sigma_2-\sigma_1)^2},
\end{eqnarray}
where we have defined $\overline{\sigma^{\alpha}}\equiv x\sigma_1^{\alpha}+(1-x)\sigma_2^{\alpha}$. As 
$\overline{\sigma^2}>\overline{\sigma}^2$ for $x\neq 0 $ or $1$, we find $\Delta>\Delta_0$. For a fixed 
parameter $r\equiv \sigma_2/\sigma_1>1$, the polydispersity coefficient as a function of $x$, $\Delta(x)$, has a 
maximum at $\displaystyle{x^*=\frac{r}{r+1}}$ with value
\begin{eqnarray}
\Delta(x^*)=\frac{r+1}{2\sqrt{r}}\sqrt{\Delta_0^2+\left(\frac{r-1}{r+1}\right)^2}.
\end{eqnarray}
For example, setting $\Delta_0=0.3$, and for $r=1.5$, 2, $2.5$ and 3, we find $\Delta(x^*)=0.368$, $0.476$, $0.579$ and $0.673$ 
respectively; we then see that the bimodality dramatically increases the effective polydispersity $\Delta$ of the mixture. 

To present the results in the following sections  
we use the packing or volume fraction $\eta_s(\gamma)$ with $s=\rm{I,N}$, which is a function 
of $\gamma$, and is defined through the zeroth moment of the distribution function 
as $\eta_s(\gamma)\equiv {\rho_s^{(0)}(\gamma)}\sigma_1^2L$. 
Specifically we will use for the presentation of results 
the values $\eta_{\rm I}\equiv \eta_{\rm I}(0)=\rho_0(0)\sigma_1^2 L$ 
and $\eta_{\rm N}\equiv \eta_{\rm N}(1)=\rho_0(1)\sigma_1^2L$, i.e. 
the cloud-I and cloud-N packing fractions. Also we use the total packing fraction of the polydisperse mixture 
$\eta(\gamma)\equiv \gamma \eta_{\rm N}(\gamma)+(1-\gamma) \eta_{\rm I}(\gamma)= \rho_0(\gamma) \sigma_1^2L$, 
the latter equality being a consequence of the lever rule. Finally we will use the length distribution functions    
corresponding to coexisting phases defined as 
$h^{(s)}(\sigma)\equiv \rho_0^{-1} \sum_{\nu}\rho_{\nu,s}(\sigma)$, ($s=\rm{I,N}$). Again, using the lever rule, 
we have $\gamma h^{(\rm{N})}(\sigma)+(1-\gamma)h^{(\rm{I})}(\sigma)=h(\sigma)$.

\subsection{Polydispersity in particle shape}
\label{shape}

In most colloidal suspensions of anisotropic particles, 
polydispersity exists not only in particle size 
but also in particle shape. As the inclusion in the theory of both types of 
polydispersities constitutes a highly nontrivial task, the usual procedure 
is to map the real particles onto effective particles of fixed shape but with an effective 
polydispersity in their sizes. In the following we describe how the effective 
polydispersity can be calculated in our particular 
system in which the main quantity that governs phase behavior is 
the particle area of the transverse section (we note again that the thickness $L$ is 
constant). 

We consider a system made of a collection of particles of 
fixed thickness $L$ and different cross sections. To be more 
precise, we suppose the latter to have the form of regular 
polygons inscribed in circumferences of different diameters $2R$ 
and also with different number of edge-lengths $n$. Thus our system 
is polydisperse in the variables $R$ and $n$, the former 
controlling the size polydispersity, while the latter 
controls the particle shape. The cross-sectional area of these 
particles can be calculated as $\displaystyle{A_n(R)=\frac{nR^2}{2}\sin\left(
\frac{2\pi}{n}\right)}$. Suppose the polydisperse coefficient 
[mean square deviation of the probability distribution function 
$h(R)$] is $\Delta$, and we define the probability 
to find a polygon with $n$ edge-lengths as $p_n$. Now we map 
our system onto an effective one, monodisperse in the number $n_0$ of 
edge-lengths, but with an effective polydispersity 
$\Delta_{\rm eff}$. We define the mapping as
\begin{eqnarray}
\langle A_{n_0}\rangle_{h_{\rm eff}}=\sum_n p_n\langle A_n\rangle_h,
\end{eqnarray}
where 
\begin{eqnarray}
{\langle A_n\rangle_{h}=\int dR h(R) A_n(R)=\frac{n\langle R\rangle^2_h\left(1+\Delta^2\right)}{2}
\sin\left(\frac{2\pi}{n}\right).}\nonumber\\
\end{eqnarray}
{We take $\langle R\rangle_h=\langle R\rangle_{h_{\rm{eff}}}$, i.e. the mean radii that follow from the 
distribution functions $h(R)$ and $h_{\rm{eff}}(R)$ are exactly the same.}
Defining the coefficient
\begin{eqnarray}
q=\frac{\sum_n p_n n\sin(2\pi/n)}{n_0\sin(2\pi/n_0)},
\end{eqnarray}
we find that the effective polydispersity coefficient can be found as
\begin{eqnarray}
\Delta_{\rm eff}=\sqrt{q(1+\Delta^2)-1}.
\end{eqnarray}  

\begin{figure}
\epsfig{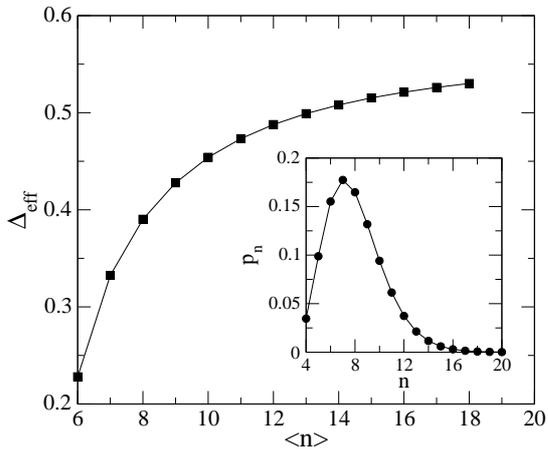}
\caption{Effective polydispersity $\Delta_{\rm eff}$ as a function 
of the mean { number of edge-lengths} 
$\langle n\rangle$ for $n_0=6$, $\Delta_0=0.3$ and $k=9$. 
Inset: probability $p_n$ as a function of $n$ for $k=9$ and $\langle n\rangle=8$.}
\label{fig1}
\end{figure}
\noindent
We have used the following expression for the probability $p_n$:
\begin{eqnarray}
p_n=\frac{(n-4+k)!}{(n-4)!k!}p^{n-4}(1-p)^{k+1},\quad n\geq 4,
\end{eqnarray}
with the triangular shape excluded, which is usually the case in experiments. 
These probabilities fulfill the normalization condition $\displaystyle\sum_{n=4}^{\infty}p_n=1$.  
Once we fix the mean 
number of edge-lengths $\langle n\rangle$, the value of $p$,  a  
function of $\langle n\rangle$ and $k$, has the form
$\displaystyle{p=\frac{\langle n\rangle-4}{\langle n\rangle +k-3}}$. 
The polydispersity {in the number of edge-lengths} can be 
{ quantified through the coefficient} 
\begin{eqnarray}
\displaystyle{\Delta_n\equiv \sqrt{\frac{\langle n^2\rangle}{
\langle n\rangle^2}-1}=\sqrt{\frac{1}{\langle n\rangle -4}+
\frac{1}{k+1 }}}.
\end{eqnarray}
Thus the number $k$ controls the polydispersity. 
In Fig. \ref{fig1} we plot the effective 
polydispersity coefficient $\Delta_{\rm eff}$ as a function of $\langle n\rangle$  
for the case $n_0=6$, $\Delta_0=0.3$ and $k=9$. As can be seen from the figure, 
$\Delta_{\rm eff}$ is a monotonically increasing function of $\langle n\rangle$ and it  
can reach values above 0.5. A particular example of $p_n$ (for $k=9$ and $\langle n\rangle=8$) is
plotted in the inset.

\section{Results}
\label{results}
\begin{figure}
\epsfig{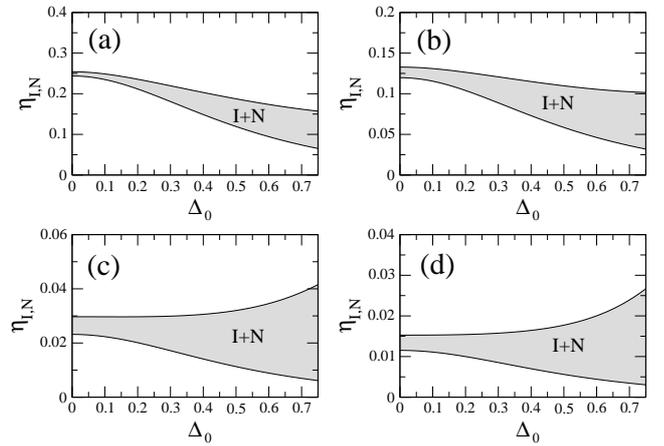}
\caption{The cloud-I and cloud-N coexisting packing fractions $\eta_{\rm I}$ and $\eta_{\rm N}$ 
as a function of polydispersity $\Delta_0$ for  
values of $\kappa$ equal to: (a) $0.2$, (b) $0.1$, (c) $0.02$ and (d) $0.01$.}
\label{fig2}
\end{figure}

We have carried out coexistence calculations following the procedure described in 
Sec. \ref{coexistence}. First we consider a polydisperse mixture with unimodal distribution 
function ($x=1$) and Gaussian tail ($q=2$), and vary the polydisperse coefficient in the range $0\leq\Delta_0\leq 0.75$. 
The results are plotted in Fig. \ref{fig2} in the $\eta_s-\Delta_0$ phase diagram for 
four different values of the aspect ratio: $\kappa=L/\sigma_1=0.2$, 0.1, 0.02, and 0.01. 

The main conclusions we can draw from these results can be summarized as follows: (i) the two-phase region 
is broadened as polydispersity is increased (in {agreement}
 with other theoretical and experimental results); 
(ii) this effect is enhanced as $\kappa$ is lowered, and (iii) the behaviour of 
$\eta_{\rm N}$ as a function of $\Delta_0$ changes with $\kappa$: for relatively high values of $\kappa$ 
it is a decreasing function, whereas for low enough $\kappa$ it becomes an increasing function for large
$\Delta_0$.  

In a recent experiment \cite{chinorri}, suspensions of polydisperse platelets were prepared from exfoliation
of pristine $\alpha$-ZrP {crystals} using TBA molecules. The resulting platelets were found to be polydisperse in 
diameter and also in shape (although most particles have approximately hexagonal geometry), with a constant thickness
equal to 26.8 \AA. Different aqueous suspensions were prepared with particles of mean aspect ratio [as measured by 
{dynamic light scattering (DLS)}] ranging from $0.001$ to $0.01$. The polydispersity coefficients of the suspensions
were estimated using Dynamic Light Scattering (DLS), and all of them were found to be in the range 18$\%$--39$\%$, 
with most samples having a value of about 30$\%$. Samples were divided into three different sets: A, B and C. Samples
corresponding to set B were obtained from nematic fractionation of an original suspension followed by dilution and, 
consequently, this set has the lowest polydispersity coefficients. The other two sets, A and C, result from the original 
synthesis and exfoliation of the pristine crystals. All platelets have negative surface charges which are partially  
neutralized by the positive charges of the TBA molecules, thus creating effective dipoles. Non-neutralized charges and dipoles 
cause the pair-interaction between two platelets to be long-ranged and repulsive. 

\begin{figure}
\epsfig{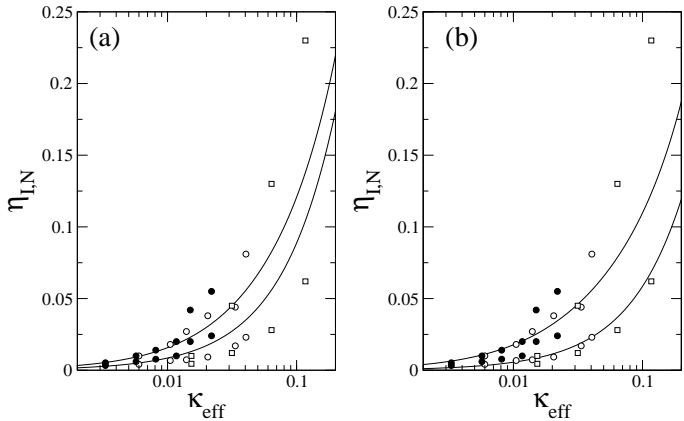}
\caption{The cloud coexisting packing fractions $\eta_{\rm{I,N}}$ as a function of 
{the aspect ratio} $\kappa_{\rm{eff}}$ for polydispersities 
fixed to (a) $\Delta_0=0.3$, and (b)  $0.5$. Open circles, filled circles and open squares correspond to 
the experimental results in Ref. \cite{chinorri} for sets A, B and C (adequately rescaled with factors $f=5$, 3 and {9},
respectively).} 
\label{fig2a}
\end{figure}

\subsection{Unimodal length distribution}
\label{uni} 
With the aim of modeling the fluid, we mapped a collection of repulsive, shape-- and diameter--polydisperse 
platelets onto effective polydisperse board-like particles interacting through excluded volume.
To properly take into account the effect of long-ranged repulsive interactions, 
the effective thickness of particles has to be larger than the thickness of the actual colloidal platelets, and
the effective aspect ratios 
$\displaystyle{\kappa_{\rm eff}\equiv \frac{L_{\rm eff}}{\langle \sigma\rangle}=\kappa \frac{L_{\rm eff}}{L}}$ are obtained 
from the real $\kappa$ by scaling by a factor $f\equiv \displaystyle{\frac{L_{\rm eff}}{L}>1}$. As polydispersity 
increases, $f$ should also increase due to the presence of platelets with large surface area which, as discussed above,
should repel each other more strongly. 

This effect is in fact obtained with the model, as shown in Fig. \ref{fig2a}, where 
the cloud packing fractions $\eta_{\rm{I,N}}$ are plotted as a function of
$\kappa_{\rm{eff}}$ for two values of polydispersity, $\Delta_0=0.3$ in (a) and $\Delta_0=0.5$ in (b), and using 
a unimodal length distribution $h(\sigma)$ with $q=2$. Also included in the figures are the
experimental results from \cite{chinorri} with $f$ set to 5, 3 and {9} for samples A, B and C, respectively.  
These values were chosen to ensure a proper agreement between theory and experiment (note that no least-square optimization
was attempted). Samples in set B are less polydisperse, and consequently $f$ is smaller. As can be seen from the figure,
these samples are relatively well described by our model using $\Delta_0=0.3$ (as in the actual samples), 
except for those experimental points with the two higher values of $\kappa_{\rm{eff}}$. However, 
for samples in set A, better agreement is obtained with $\Delta_0=0.5$. Note that polydispersity in the experimental samples 
is in diameter {\it and} also in shape which, as discussed in Sec. \ref{shape}, demands that the effective polydispersity of
a single--shaped model be higher. Finally, samples in set C, which are characterized by huge coexistence gaps, 
are not well described by unimodal length distributions. The multimodality of the distribution is probably behind
this behaviour (see following section).

\begin{figure}
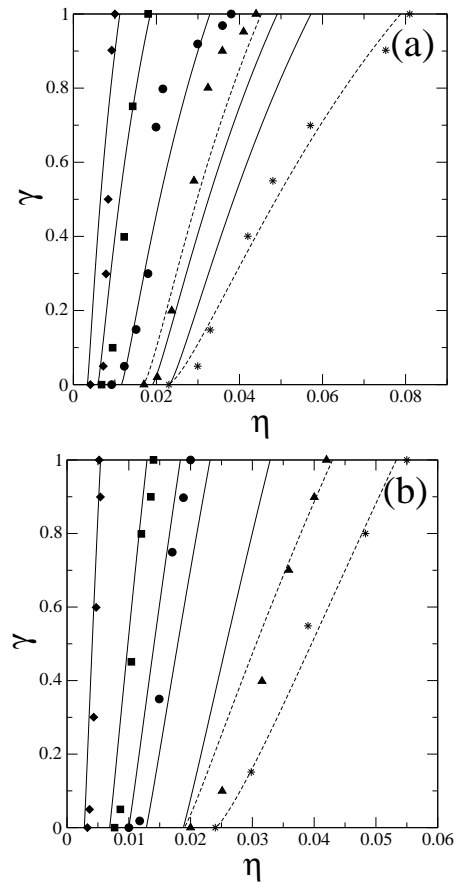

\epsfig{file=Fig4a.eps,width=2.3in}
\epsfig{file=Fig4b.eps,width=2.36in}
\caption{Percentage of total volume $\gamma$ occupied by N phase as a function of total packing fraction 
$\eta$ for values of $\kappa_{\rm{eff}}$ corresponding to the experimental sets A, panel (a), and B, panel (b).  
Rhombi, squares, circles, triangles and stars are used to show the experimental results from \cite{chinorri} as 
$\kappa_{\rm{eff}}$ is increased. Solid lines correspond to the theoretical results obtained from a unimodal 
length distribution with $q=2$ and (a) $\Delta_0=0.5$, (b) $\Delta_0=0.3$. 
Theoretical results using a bimodal distribution are represented by dashed curves.
Values of parameters that better describe experimental data represented by stars and triangles are as follows.
Panel (a): $\Delta_0=0.5$, $q=2$, $\sigma_1/L_{\rm{eff}}=10$, $\sigma_2/L_{\rm{eff}}=
17.5$ and $x=0.7$ for stars, and $\sigma_1/L_{\rm{eff}}=20$, $\sigma_2/L_{\rm{eff}}=26$ and $x=0.45$ for triangles. 
Panel (b): $\Delta_0=0.3$, $q=2$, $\sigma_1/L_{\rm{eff}}=20$, $\sigma_2/L_{\rm{eff}}=32$ and $x=0.78$ for 
stars, and $\sigma_1/L_{\rm{eff}}=20$, $\sigma_2/L_{\rm{eff}}=32.6$ and $x=0.65$ for triangles.}
\label{fig3}
\end{figure}

Fig. \ref{fig3} shows the percentage of total volume $\gamma$ occupied by the N phase as a function of the 
total packing fraction $\eta$ of the mixture for those values of $\kappa_{\rm{eff}}$ corresponding 
to the sets A [Fig. \ref{fig3}(a)] and B [Fig. \ref{fig3}(b)]. In the former case solid lines are results from 
calculations with a unimodal distribution using $q=2$ and $\Delta_0=0.5$, while in the latter case $\Delta_0=0.3$. It is
clear that when $\Delta_0=0.3$ the function $\gamma(\eta)$ is practically a linear function
[see Fig. \ref{fig3} (b)], while for $\Delta_0=0.5$ some nonlinearity is already apparent, a trend which is more pronounced 
in experiment. Coexistence gaps from theory and experiment are similar, except for the highest 
value of $\kappa_{\rm{eff}}$ in set A [stars in Fig. \ref{fig3} (a)], and for the two higher values 
of $\kappa_{\rm{eff}}$ in set B [triangles and stars in (b)]. A possible reason for these deviations will also 
be explained in the following section.   

\subsection{Bimodal distribution} 
\label{bi}
In this section we demonstrate that the multimodality of the length distribution function  
can explain both the existence of a huge I-N coexistence gap, and the strong nonlinearity of $\gamma(\eta)$.
These are two of the main features present in the experiments of \cite{chinorri} for those samples
with higher values of $\kappa_{\rm{eff}}$, i.e. samples in sets A and B. For set C
we show below that the bimodality is crucial to adequately describe experimental results.  
The origin of this behavior is the coupling between the fractionation effect, typical of polydisperse mixtures, and 
the demixing phenomenon that occurs in multicomponent mixtures of particles with sufficiently different lengths. In binary 
mixtures with very asymmetric species, entropy forces the system to segregate into two phases of very different
composition, and consequently the coexistence density gap is very large compared to that in one component systems or in 
polydisperse mixtures with unimodal size distributions.    

\begin{figure}
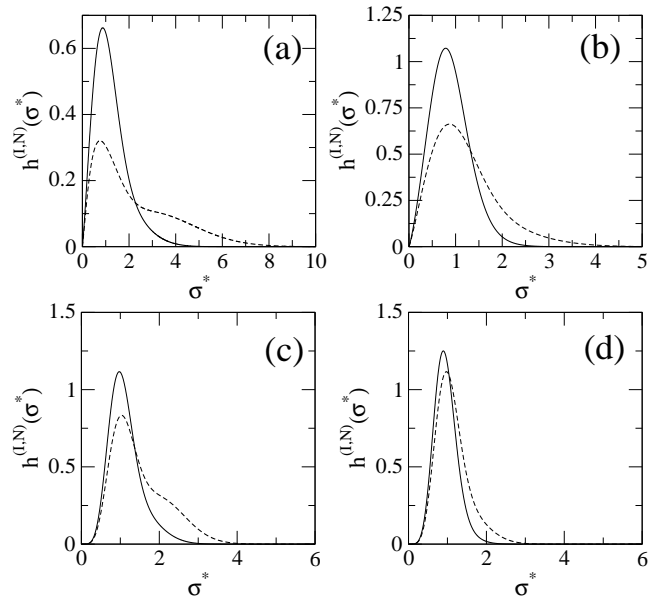

\epsfig{file=Fig5a.eps,width=3.3in}
\epsfig{file=Fig5b.eps,width=3.3in}
\caption{Length distribution functions $h^{(\rm{I,N)}}(\sigma^*)$, as a function 
of reduced length $\sigma^*=\sigma/\sigma_1$, of coexisting I (solid curves) and N (dashed curves) phases. 
Panels (a) and (c) correspond to cloud-I-shadow-N coexistence (i.e. $\gamma=0$), while (b) and (d) refer to
cloud-N-shadow-I coexistence ($\gamma=1$). The parent distribution function, which coincides with that of the 
cloud-I (for $\gamma=0$) or cloud-N (for $\gamma=1$) cases, is bimodal with the following parameters:
panels (a) and (b), $\Delta_0=0.5$, $q=2$, $\sigma_1/L_{\rm{eff}}=10$, $\sigma_2/L_{\rm{eff}}=
17.5$ and $x=0.70$; panels (c) and (d), $\Delta_0=0.3$, $q=2$,
$\sigma_1/L_{\rm{eff}}=20$, $\sigma_2/L_{\rm{eff}}=32$ and $x=0.78$.} 
\label{fig4}
\end{figure}

Dashed lines in Fig. \ref{fig3} are the theoretical predictions for $\gamma(\eta)$ 
in the case of the experimental sets A and B [triangles and stars in panels (a) and (b), respectively]. 
Calculations were based on a bimodal parent distribution function $h(\sigma)$, as described in Sec. \ref{bimodal},
and the corresponding functions for the different cases are shown in Figs. \ref{fig4} (a) and (c) (solid lines). 
Values for the parameters in $h(\sigma)$ were chosen so as to optimise agreement with the experimental data 
[stars in Figs. \ref{fig3} (a) and (b)]. As can be seen from Figs. \ref{fig4}, the parent distribution function seems to 
be unimodal in both cases, A and B, even though compositions close to 50\% were chosen ($x=0.70$ and $x=0.78$, 
respectively). The high polydispersity ($\Delta_0=0.5$ and $0.3$, respectively) creates a large overlap region 
between the two distribution functions centered at $\sigma_1$ and $\sigma_2$ [see Eqn. (\ref{h_sigma})], which results in the 
absence of a second maximum near $\sigma_2$. 

The combined effect of fractionation and demixing can also be observed in Fig. \ref{fig4} from the shape of the
distribution functions for the shadow-N phase [dashed curves in panels (a) and (c)]. Note that the fraction of platelets
with lengths $\sigma\sim \sigma_1$ is much lower than that in the coexisting (parent) I phase, but the fraction
for lengths $\sigma\geq \sigma_2$ increases dramatically. As a result, there appears a shoulder in the 
distribution function, and its decay for large $\sigma$ is much slower. The opposite effect occurs for the shadow-I 
distribution functions [solid lines in panels (b) and (d)]: now the I phase is rich in platelets with 
$\sigma\sim \sigma_1$, while the distribution function decays much faster and the 
{polydisperse} mixture is poor in large platelets.    

\begin{figure}
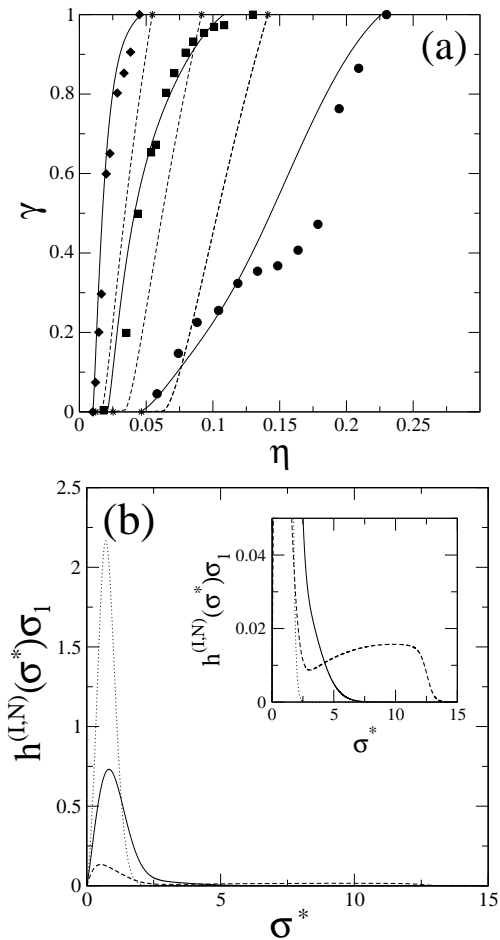

\epsfig{file=Fig6a.eps,width=2.5in}
\epsfig{file=Fig6b.eps,width=2.56in}
\caption{(a) Percentage of total volume occupied by N phase as a function of total packing fraction $\eta$ for 
values of $\kappa_{\rm{eff}}$ corresponding to the experimental samples in set C \cite{chinorri}. Rhombus, squares 
and circles are used to represent the experimental values for increasing $\kappa_{\rm{eff}}$. Dashed lines: results 
for the same value of $\kappa_{\rm{eff}}$ but using a truncated Schultz distribution function ($q=1$) with $\Delta_0=0.5$. 
Stars indicate values of $\eta_{\rm{I}}$ and $\eta_{\rm{N}}$ for each $\kappa_{\rm{eff}}$. 
Solid lines: results from a bimodal distribution functions with $q=2$ and $\Delta_0=0.5$; the other 
parameters specifying the degree of bimodality we give only for the curve that better describes the experimental data 
shown with circles. They are: $\sigma_1/L_{\rm eff}=4$, $\sigma_2/L_{\rm eff}=12$ and $x=0.9$. (b) Length distribution 
functions for the latter set of parameters. Solid, dashed and dotted lines correspond to parent, shadow-N and 
shadow-I distribution functions, respectively. Zoom of (b) shows second maximum in shadow-N distribution function.}  
\label{fig5}
\end{figure}

As mentioned above, the most dramatic disagreement between the theoretical calculations based on unimodal distributions and 
the experimental results corresponds to samples in set C (see Fig. \ref{fig2a}). 
One possible reason for this disagreement is the decay rate of the parent distribution function,
controlled by the parameter $q$. In order to check this, we have implemented our calculations using a 
truncated unimodal distribution with $q=1$ (Schultz distribution), $\Delta_0=0.5$ and the same values of $\kappa_{\rm{eff}}$  
as in Fig. \ref{fig2a}. The results, represented by means of dashed lines in Fig. \ref{fig5}(a), give a broad
coexistence gap (note that for vanishingly small values of $\gamma$, the total packing fraction 
$\eta$ rapidly increases from $\eta_{\rm{I}}$, which is shown by stars), but they fail to reproduce the experiments. 
Also, the nonlinearities of the curves $\gamma(\eta)$ are not correctly described.
 
In the same figure, the theoretical results using bimodal distribution functions with $q=2$ and $\Delta_0=0.5$ are also
plotted. The parameters $\sigma_i/L_{\rm{eff}}$ and $x$ were chosen to guarantee a reasonable agreement between theory and 
experiment so that now both the coexistence gaps and also the nonlinear behavior of $\gamma(\eta)$ are reproduced. 
Fig. \ref{fig5}(b) shows the bimodal parent and shadow-I,N coexisting length distributions 
(corresponding to those parameters which better describe the experimental points of Fig. \ref{fig5}(a) represented by circles). 
Note that the {\it bimodal} parent distribution function looks {\it unimodal}. 
Also, the shadow-N distribution exhibits a plateau in the range of scaled lengths 7--12, while 
the shadow-I is highly localized about the value 1.

\begin{figure}
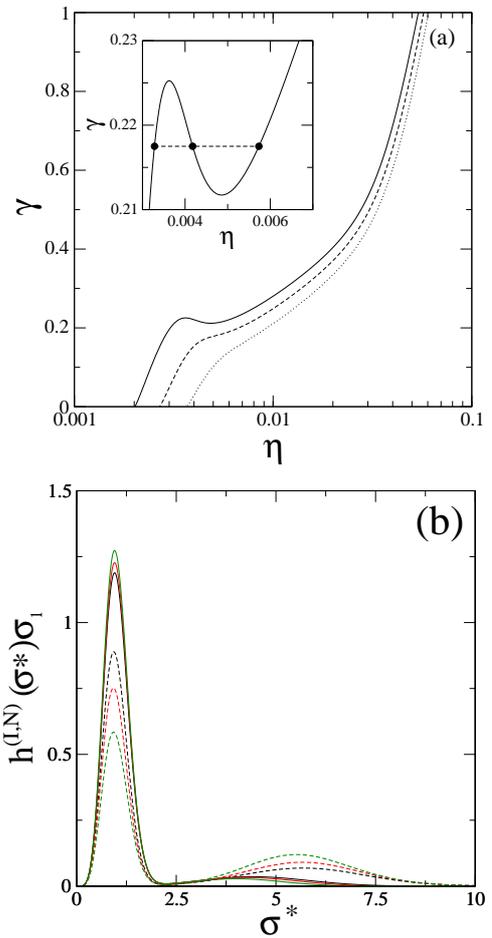

\epsfig{file=Fig7a.eps,width=2.5in}
\epsfig{file=Fig7b.eps,width=2.56in}
\caption{(a) $\gamma$ versus $\eta$ in logarithmic scale as obtained from a bimodal distribution
with $q=2$, $\Delta_0=0.3$, $x=0.85$, $\sigma_1/L_{\rm{eff}}=20$ and 
$\sigma_2/L_{\rm{eff}}=80$ (dotted curve), $90$ (dashed curve) and $100$ (solid curve). Zoom in (a) is a detail of the 
latter case showing the loop in the function $\gamma(\eta)$. (b) Three pairs of coexisting distribution 
functions $h^{(\rm{I,N})}(\sigma^*)$ corresponding to the points shown in (a) for $\gamma=0.2175$. 
Solid and dashed lines represent I and N phases, respectively, while different colors represent different values of 
$\eta$ (black, red and green in the order of increasing $\eta$).}  
\label{fig6}
\end{figure}

To illustrate how a strong nonlinearity in $\gamma(\eta)$ can emerge from the coupling between polydispersity and 
bimodality, Fig. \ref{fig6}(a) shows $\gamma$ as obtained from calculations using a bimodal distribution with 
$q=2$, $\Delta_0=0.3$, $x=0.85$, $\sigma_1/L_{\rm{eff}}=20$ and different values of $\sigma_2/L_{\rm{eff}}$. 
As the latter is increased, the nonlinearity becomes stronger up to a point where there appears a well-defined loop 
in $\gamma(\eta)$ [see inset in Fig. \ref{fig6}(a)]. We should mention here that the presence of this loop is not related 
with a triple I-N$_1$-N$_2$ or I$_1$-I$_2$-N coexistence, a fact we have proved by solving the set of equations for the 
moments and pressures of the different coexisting phases and checking that they always converge to solutions corresponding
to I-N coexistence. Fig. \ref{fig6} (b) shows the distribution functions 
$h^{(\rm{I,N})}(\sigma)$ for the three different I-N coexistences occurring for 
$\gamma=0.2175$ [symbols in inset of panel (a)] inside the loop. Clearly, as $\eta$ is increased, 
the function $h^{(\rm{N})}(\sigma)$ becomes less peaked at $\sigma_1$ but more peaked about $\sigma_2$ and with a
slower decay for large $\sigma$, while the function $h^{(\rm{I})}(\sigma)$ remains practically the same.
The presence of this loop is clearly related with the density inversion phenomenon. As can be seen in 
Fig. \ref{fig6} (b), the moments of the I and N coexisting distribution functions  
fulfill the inequalities  ${\rho_{\rm N}^{(2)}>\rho_{\rm I}^{(2)}}$ and 
${\rho_{\rm N}^{(0)}< \rho_{\rm I}^{(0)}}$ [due to the defect (excess) 
of platelets of width close to $\sigma_1$ ($\sigma_2$) in the coexisting N phase 
with respect to the I phase]. As we have defined the 
total packing fraction $\eta$ as being proportional to the zeroth moment, and this moment is lower for the N phase,
the density inversion is produced.

\begin{figure}
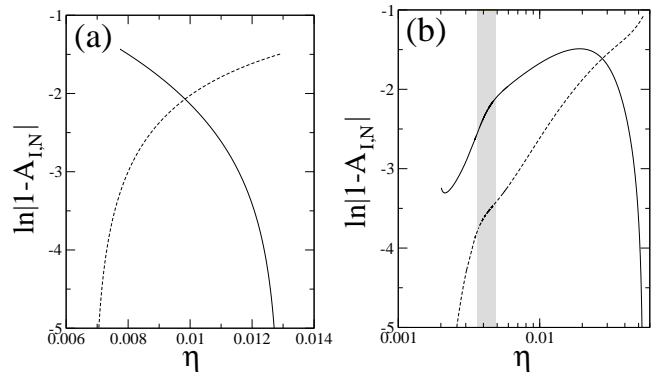

\epsfig{file=Fig8a.eps,width=1.68in}
\epsfig{file=Fig8b.eps,width=1.6in}
\caption{{ $\displaystyle{\frac{d}{d\rho_0}\ln|1-A_{\rm I,N}|}$ (dashed and 
solid lines correspond to I and N phases, respectively) as a function 
of $\eta=\rho_0 \sigma_1^2L$ for $h(\sigma)$ taken as unimodal (a) and 
bimodal (b), the latter corresponding to the results shown in Fig. \ref{fig6}. The shaded 
region in (b) shows the interval of $\eta$ where the condition (\ref{condition}) is fulfilled.}}
\label{nueva}
\end{figure}

{To better understand this behavior, we resort to 
the lever rule (\ref{const}). Dividing the whole equation by $\rho_0$ 
and integrating over $\sigma$, we find 
\begin{eqnarray}
1=\gamma(\rho_0) A_{\rm N}(\rho_0)+\left[1-\gamma(\rho_0)\right] A_{\rm I}(\rho_0),
\label{dd}
\end{eqnarray}
where we have defined $A_{\rm s}\equiv \int d\sigma h^{(s)}(\sigma)$ 
as the total area under the curve 
$h^{(s)}(\sigma)$ 
($s=\rm I,N$).  
Both these areas and $\gamma$ are functions of the parent 
number density $\rho_0$. From (\ref{dd}) we find
that the derivative of $\gamma$ with respect to $\rho_0$ is
\begin{eqnarray}
\gamma'=\frac{A_{\rm N}'(A_{\rm I}-1)+A_{\rm I}'(1-A_{\rm N})}{(A_{\rm N}-A_{\rm I})^2}.
\label{cd}
\end{eqnarray} 
In those polydisperse fluids where the parent distribution function is strictly unimodal, 
density inversion does not occur and we have 
that $A_{\rm I}\leq 1$ and $A_{\rm N}\geq 1$. Taking the 
latter inequalities into account and the fact that  
$\rho_0$ goes from $\rho_0^{(\rm I)}$ to $\rho_0^{(\rm N)}$ (the values of the parent 
number densities corresponding to the cloud-I--shadow-N coexistence, with $A_{\rm cloud-I}=1$, and 
the cloud-N--shadow I coexistence, 
with $A_{\rm cloud-N}=1$, respectively), we have most likely 
that
$A_{\rm I}'<0$ and 
$A_{\rm N}'<0$. Thus we find from Eq. (\ref{cd}) that $\gamma'>0$ in the whole  
interval $[\rho_0^{(\rm I)},\rho_0^{(\rm N)}]$. When the parent distribution function is 
bimodal, density inversion could occur. For the latter situation we have the opposite 
scenario: $A_{\rm I}\geq 1$ and $A_{\rm N}\leq 1$. Thus we could conclude that 
$A_{\rm I}'>0$ and $A_{\rm N}'>0$ and then we obtain again $\gamma'>0$ [see Eq. (\ref{cd})] 
in the whole range of $\rho_0$. 
However for strong bimodality [when 
the two peaks of $h(\sigma)$ are well visible as shown in Fig. \ref{fig6} (b)],
the sign of $A_{\rm N}'$ could change from positive to negative giving rise, 
for certain values of $\rho_0$, 
to $\gamma'<0$. 
To elucidate the conditions necessary for having a negative sign of $\gamma'$, we resort again to 
Eq. (\ref{cd}). From the latter it is easy to show that, if $\gamma'<0$, we obtain  
the condition
\begin{eqnarray}
\frac{d}{d\rho_0} \ln|1-A_{\rm N}|>\frac{d}{d\rho_0}\ln|1-A_{\rm I}|.
\label{condition}
\end{eqnarray}
In Fig. \ref{nueva} we plot the functions $\displaystyle{S(\rho_0)=\frac{d}{d\rho_0} \ln|1-A_s|}$ 
when $h(\sigma)$ is unimodal (a) and bimodal (b) (that corresponding 
to the results shown in Fig. \ref{fig6}).  
We can see that while the condition (\ref{condition}) 
is not fulfilled for any $\rho_0$ for the unimodal $h(\sigma)$, there exists, with a
bimodal $h(\sigma)$, a range of 
$\rho_0$ (shaded in the figure with grey color) for which this condition 
is fulfilled. It is easy to show, 
using Eq. (\ref{dd}), that the condition (\ref{condition}) is equivalent to 
the following inequality 
\begin{eqnarray}
\gamma |A_{\rm N}'|> (1-\gamma) A_{\rm I}'.
\label{fg}
\end{eqnarray}
Thus we conclude that $\gamma'<0$ when the rate of change in the area under the curve 
$h^{(N)}(\sigma)$ weighted with the factor $\gamma$ 
is greater than the corresponding rate of change of area of $h^{(I)}(\sigma)$ 
weighted with $1-\gamma$. 
We can see from Fig. \ref{fig6} that the distribution functions 
$h^{(N)}(\sigma)$ corresponding to the three values of $\rho_0$ shown 
($\rho_0^{(1)}<\rho_0^{(2)}<
\rho_0^{(3)}$) are dramatically different. These values
are well inside the range of $\rho_0$ where  
$A_{\rm N}'<0$ [see Fig. \ref{nueva} (b)]. 
While the first peak centered at $\sigma\sim \sigma_1$ decreases, the second peak 
centered at $\sigma\sim\sigma_2$ increases. But these changes  
do not compensate each other, resulting in the net lowering of the total area 
under the curve as $\rho_0$ increases. 
We can also see that $h^{(I)}(\sigma)$ practically remains the same for these 
three values of $\rho_0$. A small increment of the first peak with $\rho_0$ 
makes $A_{\rm I}'>0$. Finally, for
$\rho_0=\rho_0^{(2)}$ the condition (\ref{fg}) is fulfilled and we have 
at this point that $\gamma'<0$. 
}

\begin{figure}
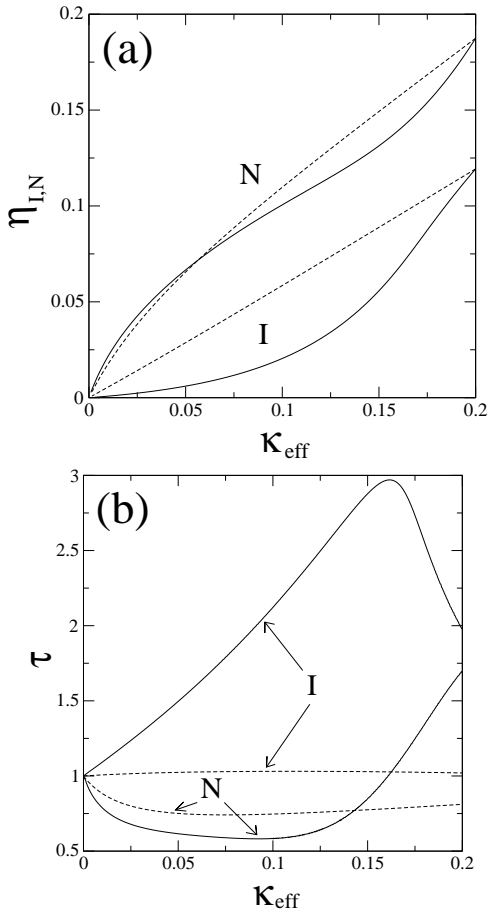

\epsfig{file=Fig9a.eps,width=2.5in}
\epsfig{file=Fig9b.eps,width=2.36in}
\caption{(a) Cloud I and N packing fractions $\eta_{\rm{I,N}}$ as a function of $\kappa_{\rm{eff}}=L_{\rm{eff}}/\sigma_1$ for 
unimodal (dashed curve) 
and bimodal (solid curve) distribution functions. 
Parameter values are set as $q=2$ and $\Delta_0=0.5$ for both distributions, while for the bimodal distribution
$x=0.85$ and $\sigma_2/\sigma_1=a+b\kappa_{\rm{eff}}$, with $a$ and $b$ chosen such that $\sigma_2/\sigma_1=1$
for $\kappa_{\rm{eff}}=0.2$ and $\sigma_2/\sigma_1=3$ for $\kappa_{\rm{eff}}=0.01$. (b) $\tau$ parameter,
calculated from $\eta_{\rm{I,N}}$ shown in (a) (see the text for definition), as a function of $\kappa_{\rm{eff}}$.}    
\label{fig7}
\end{figure}

To end this section, we show how the behaviour of the cloud coexisting packing fractions $\eta_{\rm{I,N}}$ as 
a function of $\kappa_{\rm{eff}}\equiv L_{\rm{eff}}/\sigma_1$ changes with the presence of bimodality. These
packing fractions are shown in Fig. \ref{fig7} (a) as a function of $\kappa_{\rm{eff}}$; results from unimodal and bimodal 
distribution functions are represented by dashed and solid curves, respectively. A measure of how this function behaves 
is the parameter $\displaystyle{\tau\equiv \frac{d\ln \eta_{\rm{I,N}}}{d \ln \kappa_{\rm{eff}}}}$. Note that 
if $\eta_{\rm{I,N}}\sim a\kappa_{\rm{eff}}^{\alpha}$ for small $\kappa_{\rm{eff}}$, we obtain $\tau\sim \alpha$, i.e. 
$\tau$ is a measure of the local power-law dependence of $\eta_{\rm{I,N}}$ as a function of $\kappa_{\rm{eff}}$. In 
the limit $\kappa_{\rm{eff}}\sim 0$, it can be shown that $\alpha\sim 1$ for unimodal size distributions 
and for both I and N curves, with the I curve having $\tau\gtrsim 1$ and the N curve $\tau\alt 1$ 
(the latter deviating much more from unity) [see Fig. \ref{fig7}(b)].
As polydispersity increases, this behavior is reached for lower values of $\kappa_{\rm{eff}}$. The 
parameter $\tau$ as a function of $\kappa_{\rm{eff}}$ is plotted in panel (b). From this figure we can see 
that the bimodality dramatically increases the value of $\tau$ corresponding to the cloud-I coexistence curve, while 
for the cloud-N coexistence curve the effect is the opposite for small enough $\kappa$, but to a lesser extent. 
Therefore, some caution should be exercised in extracting the power-law dependence of $\eta_{\rm{I,N}}$ with 
$\kappa_{\rm{eff}}$ by simply measuring the slope for small values of the aspect ratio. 

\section{Conclusions}
\label{conclusions}
We conclude by highlighting three results of the present work. The first concerns 
the repulsive character of pair-interactions between the colloidal platelets of Ref. \cite{chinorri}. The specific 
interactions between negatively charged platelets partially neutralized with TBA molecules (thus creating surface dipoles)
in an aqueous solvent is difficult to model. In particular, hydration layers mediate platelet interactions.
The experimental platelet volume fraction for the I-N transition is much higher than that predicted by excluded-volume-based models
for given mean aspect ratio, implying the presence of highly repulsive effective interactions between particles. It is then 
reasonable to map particles onto effective hard-core platelets with a larger effective thickness so as to 
obtain comparable results between theory and experiment. The effective thickness mainly depends 
on the diameter polydispersity (obviously platelets with large surfaces --their number depending on the 
width and the tail of the length-distribution function--, and consequently with more charges and dipoles,  
create much more repulsive effective interactions). Using this procedure, the three experimental samples, separated into 
three distinct sets A, B, and C (the major difference between them being their polydispersity) were mapped onto effective 
hard platelets with three different effective thicknesses. The hard model used was based on the restricted-orientation 
approximation for polydisperse board-like particles and a Fundamental-Measure density functional was used. 

Our theoretical results agree with experiment for sets A and B, except for samples with the highest aspect ratios. 
For set B (samples obtained via fractionation) we used a polydispersity coefficient $\Delta_0=0.3$, approximately the same value 
measured in experiments, and the theory correctly describes the experimental phase behaviour. 
For samples in set A, agreement is reached for $\Delta_0\sim 0.5$, which is higher than the experimental value. However,
we have noted (Sec. \ref{shape}) that, since colloidal particles are also polydisperse in shape, once we choose the particle 
geometry the value of the effective polydispersity should be higher. This is in fact the second important result of our work. 

The last result is related to the theoretical modeling of samples in set C. We have shown that, in this case,
a unimodal size distribution cannot describe the experimental phase behavior correctly as regards the 
huge density gap of the I-N coexistence and the strong nonlinearity in the percentage of volume 
occupied by the N phase as a function of total volume fraction. However, a bimodal size distribution 
adequately describes the gap and the nonlinearity; note that bimodality may not be apparent by direct visual 
inspection of the distribution function. 

In summary, both fractionation and demixing phenomena are important to 
explain the experimental results. These conclusions are expected to remain valid even if a more exact theoretical
treatment, including free particle orientation, could be implemented.

\appendix

\section{Free energy following the SPT}
\label{apendice}
In this section we derive, following the SPT formalism, the expression for the free-energy density of
a polydisperse Zwanzig fluid
made of hard board-like particles.
The fluid consists of a collection
of board-like particles with square polydisperse cross-section $\sigma$ and constant thickness $L$.
The main particle axes point along one of the three Cartesian axes $x,y,z$. The microscopic variables describing
the fluid are the density distribution functions $\rho_{\nu}(\sigma)$, with $\nu$ labeling the different species
($\nu=\{x,y,z\}$).

The work to insert a scaled particle of dimensions $\lambda_1\sigma$ (scaled with the width parameter
$\lambda_1$) and $\lambda_2 L$ (scaled with the thickness parameter $\lambda_2$) and pointing along the
$\nu$ direction in the polydisperse Zwanzig fluid can be calculated as
\begin{eqnarray}
&&\beta W_{\nu}(\lambda_1,\lambda_2)=-\ln\left[1-\sum_s\int d\sigma'\rho_s(\sigma')\right.\nonumber\\&&\left.\times V_{\nu}^{(\rm{excl})}
(\lambda_1\sigma,\lambda_2 L,\sigma',L)\right],
\end{eqnarray}
where the excluded volume between the particles $\nu$ and $s$ (the respective labels of particle orientations),
the latter with dimensions $\sigma'\times\sigma'\times L$, has the expression
\begin{eqnarray}
V_{\nu s}^{(\rm{excl})}(\lambda_1\sigma,\lambda_2 L,\sigma',L)&=&\prod_{\tau=\{x,y,z\}}
\left[\sigma_{\nu\tau}(\lambda_1\sigma,\lambda_2L)\right.\nonumber\\
&&\left.+\sigma_{s\tau}(\sigma',L)\right],
\end{eqnarray}
where we have defined
$\sigma_{\nu\tau}(\sigma,L)=\sigma+(L-\sigma)\delta_{\nu\tau}$, with $\delta_{\nu\tau}$ the Kronecker delta.

Following the SPT, the excess part of the chemical potential of the species $\nu$ can be 
calculated as a sum of two terms. The
first one is the second-order Taylor expansion
of $W_{\nu}(\lambda_1,\lambda_2)$ around
the point $(0,0)$ and evaluated at $(1,1)$. The second one is the product of the fluid pressure $P$ and the particle volume
$v=L\sigma^2$ (i.e. the thermodynamic work to open a cavity of dimension $v$). Thus we have
\begin{eqnarray}
\mu^{(\rm{exc})}_{\nu}(\sigma)&=&W_{\nu}(0,0)+\sum_i\frac{\partial W_{\nu}(0,0)}{\partial\lambda_i}
\nonumber\\&+&
\frac{1}{2}\sum_{i,j}\frac{\partial^2 W_{\nu}(0,0)}{\partial\lambda_i\partial\lambda_j}
+P v.
\end{eqnarray}

Taking into account the thermodynamic relations
\begin{eqnarray}
P=\sum_{\nu}\int d\sigma \mu_{\nu}(\sigma)\rho_{\nu}(\sigma)-{\cal F}/V, \quad \mu_{\nu}(\sigma)=
\frac{\delta {\cal F}/V}{\delta \rho_{\nu}(\sigma)},\nonumber\\
\end{eqnarray}
with ${\cal F}[\{\rho_{\nu}\}]$, $\mu_{\nu}$ and $V$ respectively the free-energy density functional, the chemical potential
of species $\nu$ (which is splitted in the ideal and excess part: 
$\mu_{\nu}(\sigma)=\mu^{(\rm{id})}(\sigma)+\mu_{\nu}^{(\rm{exc})}(\sigma)$), 
and the system volume, we
obtain the expression for the excess part of the free energy density 
$\Phi_{\rm exc} \equiv \beta {\cal F}_{\rm{exc}}/V$:
\begin{eqnarray}
\Phi_{\rm exc}=-n_0\ln(1-n_3)+\frac{{\bf n}_1\cdot {\bf n}_2}{1-n_3}+
\frac{n_{2x}n_{2y}n_{2z}}{(1-n_3)^2},
\label{laenergia}
\end{eqnarray}
which coincides with the uniform limit of the excess part of the free-energy density for a general
inhomogeneous fluid following the FMT \cite{Cuesta}. The weighting densities $n_{\alpha}$ are functions of
the moments $\rho_{\nu}^{(\alpha)}$ of the distribution function $\rho_{\nu}(\sigma)$:
\begin{eqnarray}
\rho_{\nu}^{(\alpha)}=\int d\sigma \rho_{\nu}(\sigma) \sigma^{\alpha},\quad \alpha=0,1,2.
\end{eqnarray}
Their expressions are
\begin{eqnarray}
&&n_0=\rho^{(0)}\equiv\sum_{\tau}\rho_{\tau}^{(0)},\quad n_3=L\rho^{(2)}\equiv L\sum_{\tau}\rho_{\tau}^{(2)},\nonumber\\
&&n_{1\nu}=L\rho_{\nu}^{(0)}+\sum_{\tau\neq \nu}\rho_{\tau}^{(1)},\quad
n_{2\nu}=\rho^{(2)}_{\nu}+L\sum_{\tau\neq \nu}\rho_{\tau}^{(1)}.\nonumber\\
\end{eqnarray}

Finally, the pressure can be calculated from (\ref{laenergia}) as
\begin{eqnarray}
\beta P=\frac{\partial\Phi_{\rm exc}}{\partial n_3}=\frac{n_0}{1-n_3}+\frac{{\bf n}_1\cdot {\bf n}_2}{(1-n_3)^2}+
\frac{2n_{2x}n_{2y}n_{2z}}{(1-n_3)^3}.\nonumber\\
\label{lapresion}
\end{eqnarray}

\begin{acknowledgments}
We thank Andres Mejia and Zhengdong Cheng for useful and illuminating discussions during the course of the present work
and for providing us with their experimental data prior to publication. We also	
acknowledge financial support from Comunidad Aut\'onoma de Madrid (Spain) under 
the R$\&$D Programme of Activities MODELICO-CM/S2009ESP-1691, and from MINECO (Spain) 
under grants {MOSAICO}, FIS2010-22047-C01 and FIS2010-22047-C04.
\end{acknowledgments}

\end{document}